\newcommand{\beq}{\begin{equation}}
\newcommand{\eeq}{\end{equation}}
\newcommand{\bea}{\begin{eqnarray}}
\newcommand{\eea}{\end{eqnarray}}
\newcommand{\bear}{\begin{eqnarray*}}
\newcommand{\eear}{\end{eqnarray*}}

\documentclass[12pt]{article}
\begin{document}

\title{New conjecture for the  $SU_q(N)$ Perk-Schultz models}
\author{F.~C.~Alcaraz$^{\rm a}$, Yu.~G.~Stroganov$^{\rm b}$\\
\small \it $^{\rm a}
$Universidade de S\~ao Paulo,
Instituto de F\'{\i}sica de S\~ao Carlos, \\
\small \it C.P. 369,13560-590, S\~ao Carlos, SP, Brazil \\
\small \it $^{\rm b}$Institute for High Energy Physics\\[-.5em]
\small \it 142284 Protvino, Moscow region, Russia}
\date{}

\maketitle


\vskip 1em


In \cite{st1} (to which we refer to hereafter by I), based on numerical 
evidence, we present a series of conjectures about the eigenspectra of 
the $SU_q(N)$ invariant Perk-Schultz Hamiltonians, given by I.1 with 
$p=0$. Subsequent extensive numerical work indicate that the conjecture (3) 
of I (I.85) is just a particular case of a more general one, relating the 
		ground-state energy of diferent $SU_q(N)$ quantum chains. This new 
		conjecture can be stated as follows:

		{\bf Conjecture:} 
The difference of the ground-state energy of the $SU_q(N)$ and $SU_{q'}(K)$ model
in an open chain of lenght $L$ is given by 

\bea \label{1}
&&	E_0[SU_q(N),  q = e^{i\pi K/(N+K)}] - 
	E_0[SU_q(N),  q = e^{i\pi N/(N+K)}] =  \nonumber \\
&&	2(1-L)\sin(\frac{ \pi (N-K)}{2(N+K)}),
\eea
where $N\neq K=1,2,\ldots$ and $E_0[SU_q(1),q)]=0$. The particular case 
$N>1$ and $K=1$ gives 
\beq \label{2}
	E_0[SU_q(N),  q = e^{i\pi /(N+1)}] = 
	2(1-L)\sin(\frac{ \pi (N-1)}{2(N+1)}),
\eeq
recovering the earlier announced conjecture (I.85).
The results (\ref{1}) and (\ref{2}) give exact 
finite-size corrections suporting  earlier 
conjectures about the operator content of the Perk-Schultz models. 
As is well known \cite{cardy} as a consequence of conformal invariance the 
 finite-size 
corrections of the ground-state energy of critical chains with free 
boundary conditions are given by 
\beq \label{3} 
E_0(L)/L = e_{\infty} + f_{\infty}/L - \frac{ \pi c}{24L^2} + o(L^{-2}),
\eeq
where $e_{\infty}$ ($f\infty$)  is the ground-state energy per site 
(surface energy) at the bulk limit 
$L\rightarrow \infty$,   
and $c$ is the conformal anomaly of the effective underlying conformal field theory 
defined on a semi-infinte plane. The relations  (\ref{3}) and 
(\ref{2})  imply that in the 
case of the $SU_q(N)$ model with $q=\exp(i\pi N/(N+1))$ the conformal anomaly 
has the value  $c=0$ for all $N\geq 2$ and 
\beq \label{4}
e_{\infty}=-f_{\infty} = -2 \sin(\frac{\pi(N-1)}{2(N+1)}.
\eeq
Moreover all the other finite-size corrections appearing in (\ref{3}) are 
identically zero! This result can be understood from the conjectured 
 operator content of the model. The conformal dimensions of the $SU_q(N)$ 
model   are  expected to be 
given by a generalized coulomb gas description 

\beq \label{5} 
x(\vec{n},\vec{m})= \frac{x_p}{2} \sum_{i,j=1}^{N-1} 
n_i C_{ij}n_j + \frac{1}{8x_p}\sum_{i,j=1}^{N-1} m_j (C^{-1})_{i,j}m_j,
\eeq
where $C$ are the $SU(N)$ Cartan matrix, and 
\beq {\label{6}}
x_p =\frac{\pi-\gamma}{2\pi},\;\;\;q = e^{i\gamma}.
\eeq
The vectors $\vec{n}=(n_1,\ldots,n_N)$ and $\vec{m}=(m_1,\ldots,m_n)$ label the 
possible values of the electric and magnetig charges in the coulomb gas 
representation. The possible values of $\vec{n}$ and 
$\vec{m}$ depend 
on the parity of the lattice and the boundary conditions where the quantum chain is 
defined. For the present  case of free boundaries the conformal anomaly of 
the effective theory is conjectured  to be given by \cite{alcmartins} 
\beq \label{7} 
c = (N-1) -12x(\vec{0}; m_1,m_2,\ldots,m_N)
\eeq
with
\beq \label{8}
m_1=m_2=\ldots = m_N= 2\frac{\gamma}{\pi}.
\eeq
In the case of relation (\ref{2}) we should use $\gamma = \pi/(N+1)$ so that  
\beq \label{9}
x_p = \frac{N}{2(N+1)}, \;\; \mbox{and} \; \; x({\vec{0};2\gamma/\pi,\dots, 
2\gamma/\pi}) = (N-1)/12, 
\eeq
and from (\ref{7}) we obtain $c=0$ for all $N$,
explaining the previous result.

The expressions (\ref{1})and (\ref{3}) for arbitrary values of $N$ and $K$ 
 also imply the interesting relations 
among the ground-state energy and surface energy in the bulk limit:
\bea \label{11}
&&e_{\infty}(SU_q(N), \gamma=\frac{\pi K}{N+K}) - 
e_{\infty}(SU_q(K), \gamma=\frac{\pi N}{N+K}) = 
-2\sin(\frac{\pi (N-K)}{N+K}),\nonumber
\eea
\bea \label{12}
&&f_{\infty}(S_qU(N), \gamma=\frac{\pi K}{N+K}) - 
f_{\infty}(S_qU(K), \gamma=\frac{\pi N}{N+K}) = 
2\sin(\frac{\pi (N-K)}{N+K}), \nonumber
\eea
as well the relation between the conformal anomalies of the 
effective conformal field theories
\beq \label{13}
c[S_qU(N), \gamma=\frac{\pi K}{N+K}]= 
c[S_qU(K), \gamma=\frac{\pi N}{N+K}]  .
\eeq
In fact this equality suport the conjectures (\ref{7}) and 
(\ref{8}) since  both sides 
of the last equation  give  the effective conformal anomaly
\beq \label{14}
c = \frac{N^2+K^2+NK-N^2K^2-K-N}{24(N+K)},
\eeq
for the related quantum chains.

\end{document}